\documentclass[twoside]{article}
\usepackage{qic,epsfig}

\newcommand{\ket}[1]{\ensuremath{\left|#1\right>}}

\begin{document}
\setlength{\textheight}{8.0truein}    

\runninghead{On the Design and Optimization of a Quantum Polynomial-Time Attack on
Elliptic Curve Cryptography  $\ldots$}
            {D. Maslov, J. Mathew, D. Cheung, and D. K. Pradhan $\ldots$}

\normalsize\textlineskip
\thispagestyle{empty}
\setcounter{page}{1}


\vspace*{0.88truein}

\alphfootnote

\fpage{1}

\centerline{\bf ON THE DESIGN AND OPTIMIZATION OF A QUANTUM}
\vspace*{0.035truein}
\centerline{\bf POLYNOMIAL-TIME ATTACK ON ELLIPTIC CURVE CRYPTOGRAPHY\footnote{An earlier version of this paper has been presented at 
the 3rd Workshop on Theory of Quantum Computation, Communication, and Cryptography, Tokyo, Japan, January 30 - February 1, 2008.}}
\vspace*{0.37truein}
\centerline{\footnotesize
DMITRI MASLOV }
\vspace*{0.015truein}
\centerline{\footnotesize\it Department of Combinatorics and Optimization, University
of Waterloo}
\baselineskip=10pt
\centerline{\footnotesize\it Waterloo, Ontario, N2L 3G1, Canada}
\vspace*{10pt}
\centerline{\footnotesize 
JIMSON MATHEW}
\vspace*{0.015truein}
\centerline{\footnotesize\it Department of Computer Science, University of Bristol}
\baselineskip=10pt
\centerline{\footnotesize\it Bristol, BS8 1UB, UK}
\vspace*{10pt}
\centerline{\footnotesize 
DONNY CHEUNG}
\vspace*{0.015truein}
\centerline{\footnotesize\it Department of Computer Science, University of Calgary}
\baselineskip=10pt
\centerline{\footnotesize\it Calgary, Alberta, T2N 1N4, Canada}
\vspace*{10pt}
\centerline{\footnotesize 
DHIRAJ K. PRADHAN}
\vspace*{0.015truein}
\centerline{\footnotesize\it Department of Computer Science, University of Bristol}
\baselineskip=10pt
\centerline{\footnotesize\it Bristol, BS8 1UB, UK}

\vspace*{0.225truein}
\publisher{(received date)}{(revised date)}

\vspace*{0.21truein}

\abstracts{
We consider a quantum polynomial-time algorithm which solves the discrete
logarithm problem for points on elliptic curves over $GF(2^m)$.  We improve
over earlier algorithms by constructing an efficient circuit for multiplying
elements of binary finite fields and by representing elliptic curve points
using a technique based on projective coordinates.  The depth of our proposed
implementation, executable in the Linear Nearest Neighbor (LNN) architecture, is $O(m^2)$, 
which is an improvement over the previous bound of $O(m^3)$ derived assuming 
no architectural restrictions. 
}{}{}

\vspace*{10pt}


\vspace*{1pt}\textlineskip    

\section{Introduction\label{sec:intro}}

Quantum computing \cite{bk:nc} has the ability to solve problems whose best
classical solutions are considered inefficient.  Perhaps the best-known example
is Shor's polynomial-time integer factorization algorithm \cite{ar:s}, where
the best known classical technique, the General Number Field Sieve, has
superpolynomial complexity $\exp{O(\sqrt[3]{n\log^2n})}$ in the number of bits
$n$ \cite{bk:gg}.  Since a hardware implementation of this algorithm on a
suitable quantum mechanical system could be used to crack the RSA cryptosystem
\cite{bk:gg}, these results force researchers to rethink the assumptions of
classical cryptography and to consider optimized circuits for the two main
parts of Shor's factorization algorithm: the quantum Fourier transform
\cite{bk:nc,co:cw} and modular exponentiation \cite{ar:mi}.  Quantum noise and
issues of scalability in quantum information processing proposals require
circuit designers to consider optimization carefully. 

Since the complexity of breaking RSA is subexponential, cryptosystems such as
Elliptic Curve Cryptography (ECC) have become increasingly popular.  The best
known classical attack on ECC requires an exponential search with complexity
$O(2^{n/2})$.  The difference is substantial: a 256-bit ECC key requires the
same effort to break as a 3072-bit RSA key.  The largest publicly broken ECC
system has a key length of 109 bits \cite{certicom}, while the key lengths of
1024 bits and higher are strongly recommended for RSA.  However, the key lengths 
represent only the communication cost of a cryptographic protocol.  It might, in 
principle, be possible that the hardware implementation of ECC were overwhelmingly 
less efficient, and then its practical efficiency would be undermined. 
However, this is not the case; indeed, the situation is rather 
opposite. Table \ref{tab:ECCRSA} compares the efficiency of CMOS circuits
(with the same clock speed) implementing 3072-bit RSA and 256-bit ECC, which both give equivalent security 
of 128 bits, in two hardware modes: optimized for space (cost), and speed (runtime). 
It is clear that the ECC implementation is more efficient. Relative efficiency 
of ECC as compared to RSA has been widely recognized. For instance, ECC has been
acknowledged by National Security Agency as a secure protocol and included in
their Suite B \cite{www:sb}. 

\begin{table}
\begin{center}
\tcaption{Comparing hardware performance of RSA-3072 and ECC-256 \cite{certicom2}.}
 \begin{tabular}{|c|c|c|}
	\hline
	Mode            & RSA-3072      & ECC-256 \\ \hline
	Space-optimized & 184 ms        & 29 ms \\
	                & 50,000 gates  & 6,660 gates \\ \hline
	Speed-optimized & 110 ms        & 1.3 ms \\
	                & 189,200 gates & 80,100 gates \\ \hline
 \end{tabular}
\label{tab:ECCRSA}
\end{center}
\end{table}

Most ECC implementations are built over $GF(2^m)$, likely, due to the efficiency of relevant hardware
and the ease of mapping a key into a binary register. Software implementations, such as ECC over 
$GF(2^{155})$, are publicly available \cite{ar:amv}, making ECC ready to use for any interested party.

There exists a quantum polynomial-time algorithm that solves 
Elliptic Curve Discrete Logarithm Problem (ECDLP) and thus cracks
elliptic curve cryptography \cite{ar:pz}. As with Shor's factorization
algorithm, this algorithm should be studied in detail by anyone interested 
in studying the threat posed by quantum computing to modern cryptography.  The quantum
algorithm for solving discrete logarithm problems in cyclic groups such as the
one used in ECC requires computing sums and products of finite field elements,
such as $GF(2^m)$ \cite{ar:j}.  Addition in $GF(2^m)$ requires only a depth-1
circuit consisting of parallel CNOT gates \cite{quant-ph/0301163}.  We present
a depth $O(m)$ multiplication circuit for $GF(2^m)$ optimized for the Linear 
Nearest Neighbor (LNN) architecture. Our circuit is based on the construction
by Mastrovito \cite{mastovito88}.  Previously, a depth $O(m^2)$ circuit in an 
unrestricted architecture was found in \cite{quant-ph/0301163}. With the use 
of our multiplication circuit the depth of the quantum discrete logarithm algorithm 
over the points on elliptic curves over $GF(2^m)$ drops from $O(m^3)$ to $O(m^2)$.
Our implementation is optimized for LNN, unlike previously constructed circuit
whose depth, if restricted to LNN, may become as high as $O(m^4)$.

The paper is organized as follows. 
In Section \ref{sec:pre} we give an overview of quantum computation, $GF(2^m)$
field arithmetic, and elliptic curve arithmetic.  Section \ref{sec:qpta}
outlines the quantum algorithm, and presents our improvements: the
$GF(2^m)$ multiplication circuit, and projective coordinate representation.
The paper concludes with some observations and suggestions for further
research.


\section{Preliminaries\label{sec:pre}}


We will be working in the quantum circuit model, where data is stored in
qubits and unitary operations are applied to various qubits at discrete time
steps as quantum gates.  We assume that any set of non-intersecting gates may
be applied within one time step.  The total number of time steps required to
execute an algorithm as a circuit is the {\em depth}.  Further details on
quantum computation in the circuit model can be found in \cite{bk:nc}.

We will make use of the CNOT, Toffoli and SWAP gates.  The CNOT gate is defined as
the unitary operator which performs the transformation
$\ket{a}\ket{b}\mapsto\ket{a}\ket{a\oplus b}$.  The Toffoli gate \cite{ar:tof}
can be described as a controlled CNOT gate, and performs the transformation
over the computational basis given by the formula
$\ket{a}\ket{b}\ket{c}\mapsto\ket{a}\ket{b}\ket{c\oplus ab}$. Finally, 
SWAP interchanges the values of the qubits, i.e., performs operation 
$\ket{a}\ket{b}\mapsto\ket{b}\ket{a}$.

\subsection{Binary Field Arithmetic}

The finite field $GF(2^m)$ consists of a set of $2^m$ elements with 
addition and multiplication operations, and additive and multiplicative
identities $0$ and $1$, respectively.  $GF(2^m)$ forms a commutative ring over
these two operations where each non-zero element has a multiplicative inverse.
The finite field $GF(2^m)$ is unique up to isomorphism.

We can represent the elements of $GF(2^m)$ where $m\geq 2$ with the help of an irreducible
{\em primitive polynomial} of the form $P(x)=\sum_{i=0}^{m-1}c_ix^i+x^m$,
where $c_{i} \in GF(2)$~\cite{pradhan78}.  The finite field $GF(2^m)$ is
isomorphic to the set of polynomials over $GF(2)$ modulo $P(x)$.  In other
words, elements of $GF(2^m)$ can be represented as polynomials over $GF(2)$ of
degree at most $m-1$, where the product of two elements is the product of
their polynomial representations, reduced modulo $P(x)$
\cite{pradhan78,hassan04}.  As the sum of two polynomials is simply the
bitwise XOR of the coefficients, it is convenient to express these polynomials
as bit vectors of length $m$.  Additional properties of finite fields can be
found in~\cite{pradhan78}.

Mastrovito has proposed an algorithm along with a classical circuit
implementation for polynomial basis (PB) multiplication
\cite{mastovito88,mastrothesis}, popularly known as the Mastrovito
multiplier.  Based on Mastrovito's algorithm, \cite{hassan04} presents a
formulation of PB multiplication and a generalized parallel-bit hardware
architecture for special types of primitive polynomials, namely trinomials,
equally spaced polynomials (ESPs), and two classes of pentanomials.

Consider the inputs $\vec{a}$ and $\vec{b}$, with
$\vec{a}=[a_0,a_1,a_2,\ldots,a_{m-1}]^T$ and
$\vec{b}=[b_0,b_1,b_2,\ldots,b_{m-1}]^T$, where the coordinates $a_i$ and
$b_i$, $0\leq i<m$, are the coefficients of two polynomials $A(x)$ and $B(x)$
representing representing two elements of $GF(2^m)$ with respect to a
primitive polynomial $P(x)$.  We use the following three matrices:
\begin{enumerate}
\item an $m\times(m-1)$ reduction matrix $M$,
\item an $m\times m$ lower triangular matrix $L$, and
\item an $(m-1)\times m$ upper triangular matrix $U$.
\end{enumerate}
Vectors $\vec{d}$ and $\vec{e}$ are defined as:
\begin{eqnarray}
\vec{d}&=&L\vec{b} \label{eqn1}\\
\vec{e}&=&U\vec{b}, \label{eqn2}
\end{eqnarray}
\noindent where $L$ and $U$ are defined as
$$L=\left[ 
\begin{array}{ccccc}
a_0     & 0       & \ldots & 0      & 0       \\ 
a_1     & a_0     & \ldots & 0      & 0       \\ 
\vdots  & \vdots  & \ddots & \vdots & \vdots  \\ 
a_{m-2} & a_{m-3} & \ldots & a_0    & 0       \\ 
a_{m-1} & a_{m-2} & \ldots & a_1    & a_0%
\end{array}%
\right],\quad 
U=\left[ 
\begin{array}{cccccc}
0      & a_{m-1} & a_{m-2} & \ldots & 0       & a_1     \\ 
0      & 0       & a_{m-1} & \ldots & 0       & a_2     \\ 
\vdots & \vdots  & \vdots  & \ddots & \vdots  & \vdots  \\ 
0      & 0       & 0       & \ldots & a_{m-1} & a_{m-2} \\ 
0      & 0       & 0       & \ldots & 0       & a_{m-1}%
\end{array}%
\right].$$
Note that $\vec{d}$ and $\vec{e}$ correspond to polynomials $D(x)$ and $E(x)$
such that $A(x)B(x)=D(x)+x^m E(x)$.  Using $P(x)$, we may construct a matrix
$M$ which converts the coefficients of any polynomial $x^mE(x)$ to the
coefficients of an equivalent polynomial modulo $P(x)$ with degree less than
$m$.  Thus, the vector
\begin{equation}\label{eqn3}
\vec{c}=\vec{d}+Q\vec{e}  
\end{equation}%
gives the coefficients of the polynomial representing the product of $\vec{a}$
and $\vec{b}$.  The construction of the matrix $M$, which is dependent on the
primitive polynomial $P(x)$, is given in \cite{hassan04}.

\subsection{Elliptic Curve Groups}

In the most general case, we define an {\em elliptic curve} over a field
$F$ as the set of points $(x,y)\in F\times F$ which satisfy the equation
$$y^2+a_1xy+a_3y=x^3+a_2x^2+a_4x+a_5.$$  By extending this curve to the
projective plane, we may include the point at infinity $\mathcal{O}$ as an
additional solution.  By defining a suitable addition operation, we may
interpret the points of an elliptic curve as an Abelian group, with
$\mathcal{O}$ as the identity element.

In the specific case of the finite field $GF(2^m)$, it is possible to reduce
the degrees of freedom in the coefficients defining the elliptic curve by the
use of linear transformations on the variables $x$ and $y$.  In addition, it
was shown in \cite{mov} that for a class of elliptic curves called {\em
supersingular} curves, it is possible to reduce the discrete logarithm
problem for the elliptic curve group to a discrete logarithm problem over a
finite field in such a way that makes such curves unsuitable for cryptography.
For $GF(2^m)$, these correspond to elliptic curves with parameter $a_1=0$.
We will restrict our attention to non-supersingular curves over $GF(2^m)$,
which are of the form $y^2+xy=x^3+ax^2+b$, where $b\neq 0$.

The set of points over an elliptic curve also forms an Abelian group with
$\mathcal{O}$ as the identity element.  For a non-supersingular curve over
$GF(2^m)$, the group operation is defined in the following manner.  Given a
point $P=(x_1,y_1)$ on the curve, we define $(-P)$ as $(x_1,x_1+y_1)$.  Given a
second point $Q=(x_2,y_2)$, where $P\neq\pm Q$, we define the sum $P+Q$ as the
point $(x_3,y_3)$ where $x_3=\lambda^2+\lambda+x_1+x_2+a$ and
$y_3=(x_1+x_3)\lambda+x_3+y_1$, with $\lambda=\frac{y_1+y_2}{x_1+x_2}$.  When
$P=Q$, we define $2P$ as the point $(x_3,y_3)$ where $x_3=\lambda^2+\lambda+a$
and $y_3=x_1^2+\lambda x_3+x_3$, with $\lambda=x_1+\frac{y_1}{x_1}$.  Also,
any group operation involving $\mathcal{O}$ simply conforms to the properties
of a group identity element.  Finally, scalar multiplication by an integer can
be easily defined in terms of repeated addition or subtraction.

The Elliptic Curve Discrete Logarithm problem (ECDLP) is defined as the
problem of retrieving a constant scalar $d$ given that $Q=dP$ for known points
$P$ and $Q$.  With this definition, we may define cryptographic protocols such 
as Diffie-Hellman or digital signature,
using the ECDLP by modifying analogous protocols using the discrete logarithm
problem over finite fields.

\section{Quantum Polynomial-Time Attack\label{sec:qpta}}

With a reversible implementation for the basic elliptic curve group
operations, it is possible to solve the ECDLP with a polynomial-depth quantum
circuit.  Given a base point $P$ and some scalar multiple $Q=dP$ on an
elliptic curve over $GF(2^m)$, Shor's algorithm for discrete logarithms
\cite{ar:s} constructs the state
$$\frac{1}{2^m}\sum_{x=0}^{2^m-1}\sum_{y=0}^{2^m-1}\ket{x}\ket{y}\ket{xP+yQ},$$
then performs a two-dimensional quantum Fourier transform over the first two
registers.  It was shown in \cite{ar:pz} that the creation of the above
state can be reduced to adding a classically known point to a
superposition of points, by using a ``double and add'' method analogous to
the square and multiply method of modular exponentiation.  Points of the
form $2^kP$ and $2^kQ$ can be classically precomputed, and then, starting
with the additive identity, group addition operations can be performed,
controlled by the appropriate bits from $\ket{x}$ or $\ket{y}$.  Note that
all of the intermediate sums must be preserved until the computation is
completed before they can be uncomputed.

\begin{figure}[tb]
\centerline{\includegraphics[height=50mm]{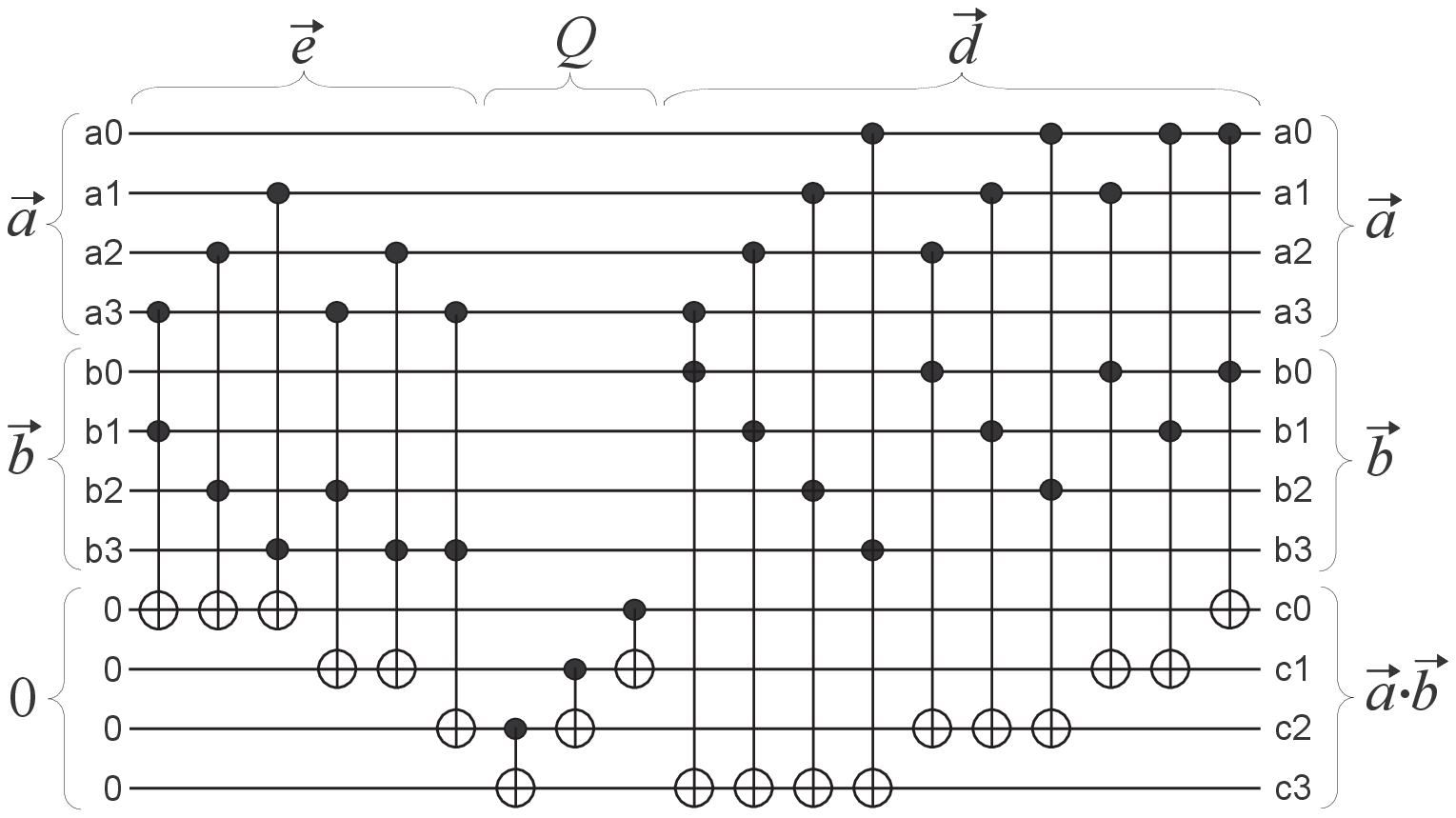}}
\vspace*{13pt}
\fcaption{\label{gf16_mult_ex}Circuit for $GF(2^{4}$) multiplier with $P(x) = x^{4}+x+1$.}
\end{figure}

\subsection{Linear Depth Circuit for $GF(2^m)$ Multiplication in the LNN Architecture \label{sec:gfm}}

We now discuss how to implement multiplication over $GF(2^m)$ as a quantum
circuit.  Firstly, using equations (\ref{eqn1}--\ref{eqn3}), derive expressions for
$\vec{d}$, $\vec{e}$ and $\vec{c}$.  We next perform the following steps breaking 
the entire computation into three distinct stages/circuits:
\begin{enumerate}
\item Compute $\vec{e}$ in an ancillary register of $m$ qubits.
\item Transform $\vec{e}$ into $M\vec{e}$, using a linear reversible
implementation.
\item Compute and add $\vec{d}$ to the register occupied by $M\vec{e}$.
\end{enumerate}
We illustrate the above steps with an example using $P(x)=x^4+x+1$.
Expressions for $\vec{d}$ and $\vec{e}$ derived from equations
(\ref{eqn1}--\ref{eqn2}) are shown below.
$$\vec{d}=\left[
\begin{array}{c}
a_{0}b_{0} \\
a_{1}b_{0}+a_{0}b_{1} \\
a_{2}b_{0}+a_{1}b_{1}+a_{0}b_{2} \\
a_{3}b_{0}+a_{2}b_{1}+a_{1}b_{2}+a_{0}b_{3}%
\end{array}\right], \quad
\vec{e}=\left[
\begin{array}{c}
a_{3}b_{1}+a_{2}b_{2}+a_{1}b_{3} \\
a_{3}b_{2}+a_{2}b_{3} \\
a_{3}b_{3}%
\end{array}
\right].$$
We also construct the matrix $M=\left[ 
\begin{array}{ccc}
1 & 0 & 0 \\
1 & 1 & 0 \\
0 & 1 & 1 \\
0 & 0 & 1 
\end{array}
\right].$

\noindent From (\ref{eqn3}), we compute the multiplier output
$\vec{c}=\vec{d}+M\vec{e}=\left[
\begin{array}{c}
d_{0} + e_{0} \\
d_{1} + e_{1} + e_{0} \\
d_{2} + e_{1} + e_{2} \\
d_{3} + e_{2}%
\end{array}\right].$

\begin{enumerate} 
\item We first compute $e_{0}$, $e_{1}$, and $e_{2}$ in the ancilla, as shown
in Figure \ref{gf16_mult_ex} (gates 1--6).

\item We next implement the matrix transformation $M\vec{e}$ (gates 7--9).

\item Finally, we compute the coefficients $d_i$, $0\leq i<m$, and add them to
the ancilla to compute $\vec{c}$ (gates 10--19).
\end{enumerate}


At this point, we have a classical reversible circuit which implements the
transformation $\ket{a}\ket{b}\ket{0}\mapsto\ket{a}\ket{b}\ket{a\cdot b}$.
However, if we input a superposition of field elements, then the output
register will be entangled with the input.  If one of the inputs, such as
$\ket{b}$ is classically known, then we may also obtain $\ket{b^{-1}}$
classically.  Since we may construct a circuit which maps
$\ket{a\cdot b}\ket{b^{-1}}\ket{0}\mapsto\ket{a\cdot b}\ket{b^{-1}}\ket{a}$,
we may apply the inverse of this circuit to the output of the first circuit to
obtain $\ket{a}\ket{b}\ket{a\cdot b}\mapsto\ket{0}\ket{b}\ket{a\cdot b}$ using
an ancilla set to $\ket{b^{-1}}$.  This gives us a quantum circuit which takes
a quantum input $\ket{a}$ and classical input $\ket{b}$, and outputs
$\ket{a\cdot b}\ket{b}$.  When $\ket{b}$ is not a classical input,
the output of the circuit may remain entangled with the input, and other
techniques may be required to remove this entanglement.  However, we emphasize
that this is not required for a polynomial-time quantum algorithm for the
ECDLP \cite{ar:pz}.

In some circumstances, we may derive exact expressions for the number of gates
required in the GF multiplication circuit.
\begin{lemma}\label{lem:1}
A binary field multiplier for primitive polynomial $P(x)$ can be designed
using at most $2m^2-1$ gates.  If $P(x)$ is a trinomial or the all-one
polynomial, where each coefficient is $1$, we require only $m^2+m-1$ gates.
\end{lemma}

{\em Proof.}
There are three phases to the computation: computing $\vec{e}$, computing
$M\vec{e}$, and adding $\vec{d}$ to the result.  For $\vec{e}$ and $\vec{d}$,
each pair of coefficients which are multiplied and then added to another qubit
requires one Toffoli gate.  This requires
$$\sum_{i=0}^{m-1}i=\frac{m(m-1)}{2},\mbox{ and }
\sum_{i=0}^{m}i=\frac{m(m+1)}{2}$$
gates respectively, for a total of $m^2$ gates.  Next, consider the
implementation of the transformation $M$.

In general, $m^2-1$ CNOT gates suffice for any linear reversible computation
defined by the matrix $M$ in equation (\ref{eqn3}) \cite{quant-ph/0703211}.
This gives a general upper bound of $2m^2-1$ gates.  In the specific case of
the All-One-Polynomial, the operation $M$ consists of adding $e_1$ to each of
the other qubits, requiring $m-1$ CNOT operations.  This gives a total of
$m^2+m-1$ operations.

For a trinomial, we have a primitive polynomial $P(x)=x^m+x^k+1$ for some
constant $k$ such that $1\leq k<m$.  To upper bound the number of gates required to
implement $M$, we may consider the inverse operation, in which we begin with a
polynomial of degree at most $m-1$, and we wish to find an equivalent
polynomial where each term has degree between $m-1$ and $2m-2$.  Increasing
the minimum degree of a polynomial requires one CNOT operation, and this must
be done $m-1$ times.  Again, this gives a total of $m^2+m-1$ operations. {\em QED}

\subsubsection{Parallelization}

\begin{figure}[t]
\centerline{\includegraphics[height=92mm]{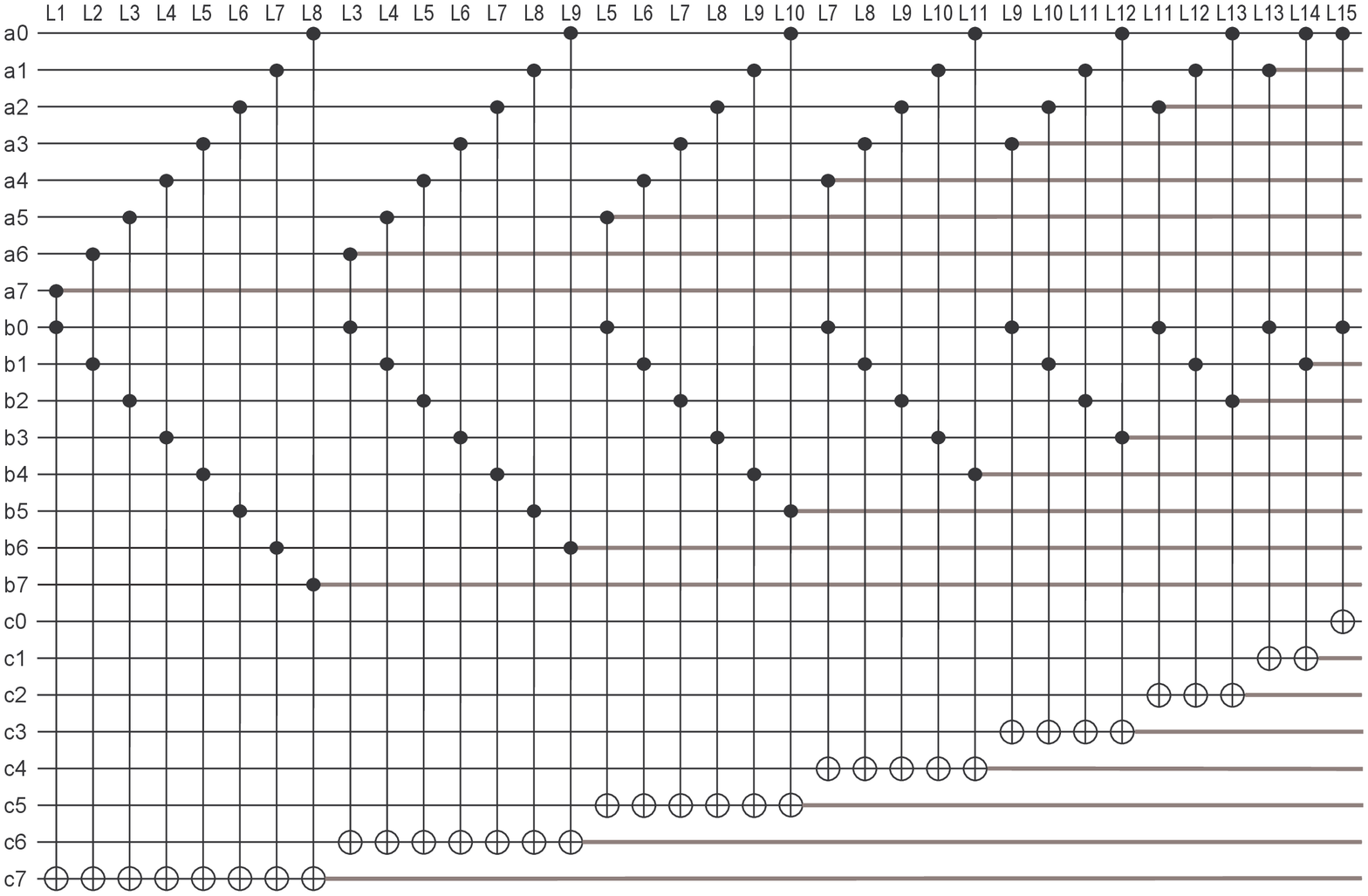}}
\vspace*{13pt}
\fcaption{\label{8bit}Subcircuit computing $\vec{d}$ illustrated in the case of multiplication in $GF(2^8)$. 
$L*$ indicate which stage does the given gate gets executed at. The qubit line 
turns gray after a given qubit was used last during the computation.} 
\end{figure}

We construct a parallelized version of this network by considering the three
parts of the computation: computation of $\vec{e}$, multiplication by $M$ and in-place computation of $\vec{d}$.
For $\vec{e}$ and $\vec{d}$, note that given coefficients $a_i$ and $b_j$
where the value of $i-j$ is fixed, the target qubit of each separate term
$a_ib_j$ is different.  This means that they may be performed in parallel.  In
the case of $\vec{e}$, we evaluate $a_ib_j$ whenever $i+j\geq m$.  This means
that the values of $i-j$ may range from $-(m-2)$ to $m-2$, giving a depth
$2m-3$ circuit for finding $e$.  Similarly, for $\vec{d}$, we evaluate
$a_ib_j$ whenever $i+j<m$.  The values of $i-j$ range from $-(m-1)$ to $m-1$,
giving a depth $2m-1$ circuit.  Evaluation of $\vec{d}$ in the case of $GF(2^8)$ 
is illustrated in Figure \ref{8bit}.

In \cite{quant-ph/0701194}, it is shown that every linear computation, 
such as that of the product $M\vec{e}$, can be done in a
linear number of stages, with a depth of at most $5m$.  Thus, a total
depth of $(2m-3)+5m+(2m-1)=9m+O(1)$ suffices to implement the multiplication circuit.  
As such, an implementation which replaces the Toffoli gate with 2-qubit gates 
(\cite{bk:nc}, page 182) can be done by a circuit with the depth upper 
bounded by the expression $25m+O(1)$.

\subsubsection{Execution of the multiplication circuit in LNN}

\begin{figure}[t]
\centerline{\includegraphics[height=100mm]{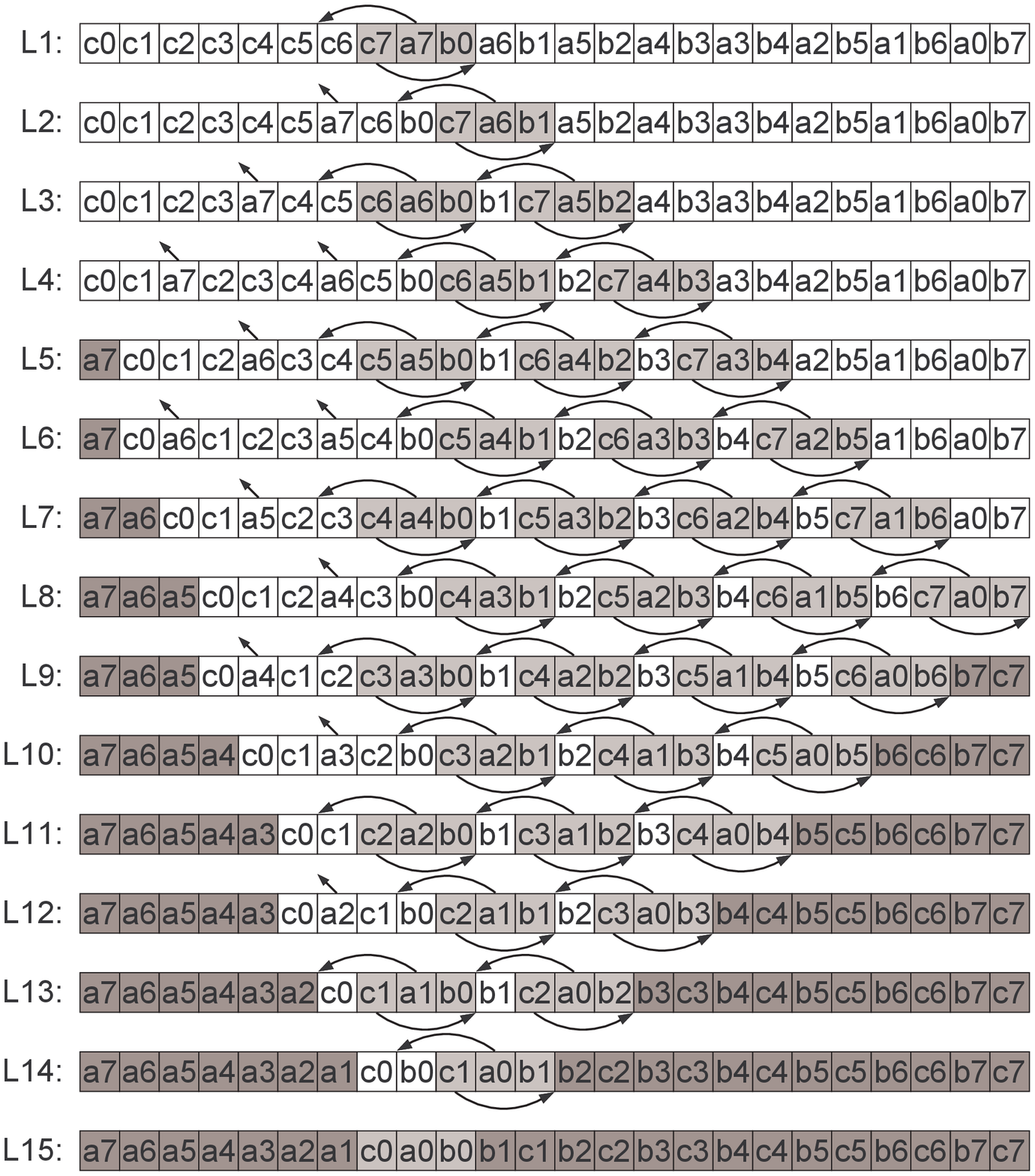}}
\vspace*{13pt}
\fcaption{\label{GFmultinLNN}Qubit permutation stages for adaptation of the GF multiplication 
circuit the the LNN architecture, illustrated in the case $GF(2^8)$ and the computation
of $\vec{d}$.  The arrows indicate which position do the individual qubits get swapped to. 
In particular, arrow $\nwarrow$ indicates that the given qubit will not be used in the 
remainder of the computation, and it gets moved to the rightmost position.  Triples of 
qubits highlighted light gray experience application of the Toffoli gates.  Dark grey
color is used to highlight qubits that are not used in the remainder of computation.}
\end{figure}

In this subsection we explain how to execute the GF multiplication circuit in linear 
depth $O(m)$ in the LNN architecture. Our circuit consists of three distinct stages: 
creation of $\vec{e}$, followed by 
a linear reversible transformation, and the in-place calculation of $\vec{d}$. 
As shown in \cite{quant-ph/0701194}, the middle part of this calculation can be 
executed as a depth $5m$ computation in the LNN architecture. We next show that first and third 
parts in our construction can also be modified to become a linear depth computation 
in the LNN. At this point, we note that both subcircuits share identical structure, 
and as such we only need to consider either one. We choose the circuit for computing 
$\vec{d}$. In the following, we will use its parallelized version described in 
the previous subsection and separate every two computational stages by a depth-2 qubit 
swapping stage such as to make it possible to execute the entire computation in the LNN in 
linear depth. 

First, prepare the qubits in the LNN connectivity pattern $c_{0}-c_{1}-...-c_{m-1}
-a_{m-1}-b_{0}-a_{m-2}-b_{1}-...-a_{0}-b_{m-1}$. For that, at most linear depth 
qubit swapping stage is required, no matter what is the starting connectivity pattern.
Next, execute a computational stage. For every Toffoli gate $TOF(c_i;a_j,b_k)$ applied
and a qubit $x$ on the left from $c_i$ in the present LNN connectivity pattern use the 
depth-2 swapping stage SWAP$(c_i,a_j)\;$SWAP$(x,a_j)\;$SWAP$(c_i,b_k)$ to prepare the qubits 
for the next computational stage (LNN connectivity pattern $x-c_i-a_j-b_k$ gets transformed to $a_j-x-b_k-c_i$). 
The workings of such adaptation to the LNN architecture are illustrated 
in Figures \ref{8bit} and \ref{GFmultinLNN}. Note that the number of swapping stages 
is no more than twice the number of the computational stages. Therefore, the entire 
computation can be executed in linear depth, not exceeding $34m+O(1)$ (counting 
1- and 2-qubit operations), in the LNN architecture.

\subsection{Projective Representation}

When points on an elliptic curve are represented as {\em affine coordinates}
$(x,y)$, performing group operations on such points requires finding the
multiplicative inverse of elements of $GF(2^m)$.  This operation takes much
longer to perform than the other field operations required, and it is
desirable to minimize the number of division operations.  For
example, \cite{quant-ph/0407095} gives a quantum circuit of depth $O(m^2)$
which uses the extended Euclidean algorithm.

By using {\em projective} coordinate representation, we can perform group
operations without division.  Instead of using two elements of $GF(2^m)$ to
represent a point, we use three elements, $(X,Y,Z)$ to represent the point
$(\frac{X}{Z},\frac{Y}{Z})$ in affine coordinates.  Dividing $X$ and $Y$ by a
certain quantity is now equivalent to multiplying the third coordinate $(Z)$
by this quantity.  Extensions to this concept have also been explored, where
different information about an elliptic curve point is stored in several
coordinates.  Another advantage to projective coordinates is that the point at
infinity $\mathcal{O}$ can simply be represented by setting $Z$ to zero.
However, in order to retrieve the elliptic curve point in the affine
representation, we still need to perform one multiplicative inversion at the
end.

To represent the point $(X,Y)$, we simply begin with the representation
$$\ket{P(X,Y)}=\ket{X}\ket{Y}\ket{1}.$$  As we perform elliptic curve group
operations, the third coordinate will not remain constant.

Exact formulas for point addition in projective coordinates can be easily
derived by taking the formulas for the affine coordinates under a common
denominator and multiplying the $Z$ coordinate by this denominator.  These
are detailed in \cite{ar:hlm}.  Since the ECDLP can be solved by implementing
elliptic curve point addition where one point is ``classically known''
\cite{ar:pz}, we may implement these formulas using the multiplication
algorithm presented in Section \ref{sec:gfm} and by being careful to uncompute
any temporary registers used.  Since the number of multiplication operations
used in these formulas is fixed, we may implement elliptic curve point
addition with a known classical point with a linear depth circuit.

Finally, to construct the state required for solving the ECDLP, we use the
standard ``double and add'' technique, which requires implementing the point
addition circuit for each value $2^iP$ and $2^iQ$, where $0\leq i<m$.  Note
that these points are classically known, so that at each step, we are
performing point addition with one classically known point.  When the final
state
$$\frac{1}{2^m}\sum_{x=0}^{2^m-1}\sum_{y=0}^{2^m-1}\ket{x}\ket{y}\ket{xP+yQ}$$
is constructed, each $\ket{xP+yQ}$ will consist of three coordinates
$\ket{X}\ket{Y}\ket{Z}$.  Since the presence of the third coordinate $Z$ will
interfere with the discrete logarithm algorithm, we must revert to an affine
coordinate representation.  An algorithm to compute the multiplicative inverse
of an element of $GF(2^m)$ using an $O(m^2)$-depth circuit is given in
\cite{quant-ph/0407095}.  Using $\ket{Z^{-1}}$, we may compute
$\ket{XZ^{-1}}\ket{YZ^{-1}}$, as required, before uncomputing $\ket{Z^{-1}}$.
Since $\ket{X}\ket{Y}\ket{Z}$ must now be uncomputed, this step must occur
before any of the temporary registers used in computing them are themselves
uncomputed.  The result is the desired state
$$\frac{1}{2^m}\sum_{x=0}^{2^m-1}\sum_{y=0}^{2^m-1}\ket{x}\ket{y}\ket{xP+yQ}$$
in affine coordinates.
As a final detail, we also need to address the {\em point at infinity}, $\mathcal{O}$, which is the identity element of the elliptic curve group.
In projective coordinates, $\mathcal{O}$ is represented by any $(X,Y,Z)$ where $Z=0$.
In this case, we will not be able to perform multiplicative inversion on $Z$.
However, since the ensuing quantum Fourier transform only requires that each point have a consistent representation, we may simply select the coordinates 
of a point which is known not to lie on the elliptic curve to represent $\mathcal{O}$.
The final registers can simply be set to these coordinates in the case that $Z=0$.

This represents an improvement on the algorithm of \cite{quant-ph/0407095},
as multiplicative inversion is used only once, at the end of this algorithm,
rather than at each elliptic curve point operation.  In total, we perform
$O(m)$ instances of the linear depth multiplication circuit, one instance of
the $O(m^2)$-depth multiplicative inversion circuit, and finally, a quantum
Fourier transform.  This gives a final depth complexity of $O(m^2)$ for the
circuit which solves the ECDLP over $GF(2^m)$ in the LNN architecture.  This
improves the previously known upper bound of $O(m^3)$ \cite{ar:pz}.

\section{Conclusion\label{sec:c}}

We considered the optimization of the quantum attack on the elliptic curve
discrete logarithm problem, on which elliptic curve cryptography is based. Our
constructions include a linear depth circuit for binary field multiplication
and efficient data representation using projective coordinates.  Our main
result is the depth $O(m^2)$ circuit executable in the LNN architecture 
for computing the discrete logarithm over
elliptic curves over $GF(2^m)$.  Further research may be devoted toward a
better optimization, further study of architectural implications, and the fault
tolerance issues.

Interestingly, our circuit is slightly 
(by a logarithmic factor) more efficient than the best known 
circuit for integer factoring optimized for the LNN 
architecture, allowing linear ancilla and assuming gates with diminishingly small parameters cannot 
be used \cite{kutin}. (We believe this is related to necessity of performing 
carry over during the integer addition, while it is not required for 
the addition over $GF(2^m)$.) However, our circuit reduces an exponential classical
search to a polynomial time quantum, whereas integer factoring can be done 
classically with a subexponential time algorithm. Considering relative 
efficiency of ECC as compared to RSA, we suggest referring to the ability 
to solve ECDLP as a stronger practical argument for quantum computing.

\nonumsection{References}


\begin{thebibliography}{000}
\bibitem{quant-ph/0703211}
To see that this is true, write a reversible linear function in the ($m\times m$) matrix form, 
and apply Gauss-Jordan elimination algorithm to synthesize a circuit using CNOT gates.
This requires application of at most $m^2$ gates, one per each entry of the matrix; 
and it may be observed that at least one gate needs not be applied.
The resulting gate count is thus at most $m^2-1$.
\end{thebibliography}
\end{document}